\documentclass[twoside,twocolumn,9pt]{article}
\usepackage{extsizes}
\usepackage[super,sort&compress,comma]{natbib} 
\usepackage[version=3]{mhchem}
\usepackage[left=1.5cm, right=1.5cm, top=1.785cm, bottom=2.0cm]{geometry}
\usepackage{balance}
\usepackage{mathptmx}
\usepackage{sectsty}
\usepackage{graphicx} 
\usepackage{lastpage}
\usepackage[format=plain,justification=justified,singlelinecheck=false,font={stretch=1.125,small,sf},labelfont=bf,labelsep=space]{caption}
\usepackage{float}
\usepackage{fancyhdr}
\usepackage{fnpos}
\usepackage[english]{babel}
\usepackage{array}
\usepackage{droidsans}
\usepackage{charter}
\usepackage[T1]{fontenc}
\usepackage[usenames,dvipsnames]{xcolor}
\usepackage{setspace}
\usepackage[compact]{titlesec}
\usepackage{hyperref}
\usepackage{tabularx}
\usepackage{lipsum}
\usepackage{comment}
\usepackage{listings}
\usepackage{amssymb}

\usepackage{enumitem}
\usepackage{epstopdf}

\usepackage[dvipsnames]{xcolor}
\usepackage[normalem]{ulem}

\definecolor{cream}{RGB}{222,217,201}

\graphicspath{ {./images/} }

\begin{document}
	
	\pagestyle{fancy}
	\thispagestyle{plain}
	\fancypagestyle{plain}{
		\renewcommand{\headrulewidth}{0pt}
	}
	
	\makeFNbottom
	\makeatletter
	\renewcommand\LARGE{\@setfontsize\LARGE{15pt}{17}}
	\renewcommand\Large{\@setfontsize\Large{12pt}{14}}
	\renewcommand\large{\@setfontsize\large{10pt}{12}}
	\renewcommand\footnotesize{\@setfontsize\footnotesize{7pt}{10}}
	\makeatother
	
	\renewcommand{\thefootnote}{\fnsymbol{footnote}}
	\renewcommand\footnoterule{\vspace*{1pt}%
		\color{cream}\hrule width 3.5in height 0.4pt \color{black}\vspace*{5pt}} 
	\setcounter{secnumdepth}{5}
	
	\makeatletter 
	\renewcommand\@biblabel[1]{#1}            
	\renewcommand\@makefntext[1]%
	{\noindent\makebox[0pt][r]{\@thefnmark\,}#1}
	\makeatother 
	\renewcommand{\figurename}{\small{Fig.}~}
	\sectionfont{\sffamily\Large}
	\subsectionfont{\normalsize}
	\subsubsectionfont{\bf}
	\setstretch{1.125} 
	\setlength{\skip\footins}{0.8cm}
	\setlength{\footnotesep}{0.25cm}
	\setlength{\jot}{10pt}
	\titlespacing*{\section}{0pt}{4pt}{4pt}
	\titlespacing*{\subsection}{0pt}{15pt}{1pt}
	
	\fancyfoot{}
	\fancyfoot[LO,RE]{\vspace{-7.1pt}\includegraphics[height=9pt]{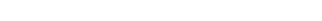}}
	\fancyfoot[CO]{\vspace{-7.1pt}\hspace{13.2cm}\includegraphics{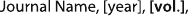}}
	\fancyfoot[CE]{\vspace{-7.2pt}\hspace{-14.2cm}\includegraphics{head_foot/RF}}
	\fancyfoot[RO]{\footnotesize{\sffamily{1--\pageref{LastPage} ~\textbar  \hspace{2pt}\thepage}}}
	\fancyfoot[LE]{\footnotesize{\sffamily{\thepage~\textbar\hspace{3.45cm} 1--\pageref{LastPage}}}}
	\fancyhead{}
	\renewcommand{\headrulewidth}{0pt} 
	\renewcommand{\footrulewidth}{0pt}
	\setlength{\arrayrulewidth}{1pt}
	\setlength{\columnsep}{6.5mm}
	\setlength\bibsep{1pt}
	
	\makeatletter 
	\newlength{\figrulesep} 
	\setlength{\figrulesep}{0.5\textfloatsep} 
	
	\newcommand{\topfigrule}{\vspace*{-1pt}%
		\noindent{\color{cream}\rule[-\figrulesep]{\columnwidth}{1.5pt}} }
	
	\newcommand{\botfigrule}{\vspace*{-2pt}%
		\noindent{\color{cream}\rule[\figrulesep]{\columnwidth}{1.5pt}} }
	
	\newcommand{\dblfigrule}{\vspace*{-1pt}%
		\noindent{\color{cream}\rule[-\figrulesep]{\textwidth}{1.5pt}} }
	
	\makeatother
	
	\twocolumn[
	\begin{@twocolumnfalse}
		{\includegraphics[height=30pt]{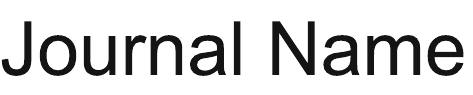}\hfill\raisebox{0pt}[0pt][0pt]{\includegraphics[height=55pt]{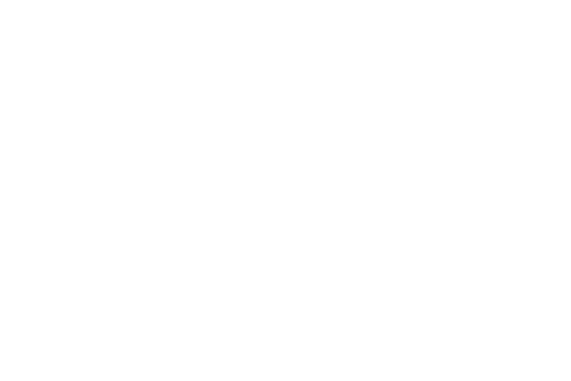}}\\[1ex]
			\includegraphics[width=18.5cm]{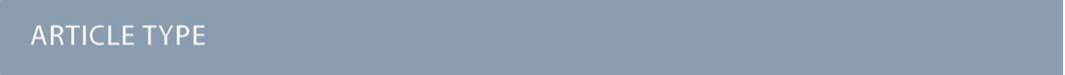}}\par
		\vspace{1em}
		\sffamily
		\begin{tabular}{m{4.5cm} p{13.5cm} }
			
			\includegraphics{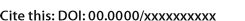} & \noindent\LARGE{\textbf{Growth and Transport Properties of InAsSb Nanoflags$^\dag$}} \\
			\vspace{0.3cm} & \vspace{0.3cm} \\
			
			& \noindent\large{Sebastian Serra,\textit{$^{a}$} Gaurav Shukla,\textit{$^{a}$} Giada Bucci,\textit{$^{a}$} Robert Sorodoc,\textit{$^{a}$} Valentina Zannier,\textit{$^{a}$} Fabio Beltram,\textit{$^{a}$} Lucia Sorba,\textit{$^{a}$}} and  Stefan Heun\textit{$^{a}$}\\
			
			\includegraphics{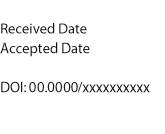} & \noindent\normalsize{The present work reports, for the first time, the growth of high-quality free-standing InAsSb nano\-flags and their electronic properties. Different growth conditions have been explored, and zinc-blende InAsSb nanoflags of various composition have been obtained. In particular, InAs$_{0.77}$Sb$_{0.23}$ nano\-flags are on average $(2000 \pm 180)$~nm long, $(640 \pm 50)$~nm wide, and $(130 \pm 30)$~nm thick. We show that these nano\-flags have a Landé $g$-factor larger than InAs and InSb and a mobility comparable to those of the best performing InAs and InSb nanoflags. Besides, we show evidence for a surface Fermi level pinning in the conductance band of these InAs$_{0.77}$Sb$_{0.23}$ nano\-flags, similar to the well-known behavior of InAs. This promises to make InAsSb easy to couple to superconductors, while keeping or improving many of the features that make InSb an interesting material for quantum applications.} \\
			
		\end{tabular}
		
	\end{@twocolumnfalse} \vspace{0.6cm}
	
	]
	
	\renewcommand*\rmdefault{bch}\normalfont\upshape
	\rmfamily
	\section*{}
	\vspace{-1cm}

	
	\footnotetext{\textit{$^{a}$~NEST, Istituto Nanoscienze-CNR and Scuola Normale Superiore, 56127 Pisa, Italy. Fax: +39-050-509-550; Tel: +39-050-509-118; E-mail: lucia.sorba@nano.cnr.it; stefan.heun@nano.cnr.it}}
	
	\footnotetext{\dag~Supplementary Information available: [growth geometry, device geometry, absence of current pinch-off, two channel conduction model, Landau fan diagram, effective mass]. See DOI: 00.0000/00000000.}
	

	

\section{Introduction}

III-V semiconductors have been the subject of extensive study, as they are considered ideal for a variety of electronic and optoelectronic applications, like high-performance transistors,\cite{III-V:transistors, III-V:electronics, InAs:high_performance_transistors} light emitters,\cite{InAs:light-emitting_NWs, InAsP/InP:QD_NWs_light_emission} photodetectors,\cite{III-V:photodetectors, InAs:photodetectors_UV-IR} and photovoltaics.\cite{III-V:solar_cells, III-V:NW_solar_cells} In particular, high quality crystals with narrow band gap, strong spin-orbit interaction, and large Landé $g$-factor are considered promising platforms for optoelectronics, spintronics, and quantum computing applications. Among these compounds, indium arsenide (InAs), indium antimonide (InSb), and the ternary alloy indium-arsenide-antimonide (InAsSb), represent ideal candidates for these purposes.

InAs has a narrow and direct bandgap ($\sim 0.39$~eV in wurtzite structure and $\sim 0.35$~eV in zincblende structure, at a temperature of $300$~K),\cite{InAs:WZ_ZB_differences} low effective mass ($m^*_{\text{InAs}} = 0.023\, m_e$, where $m_e$ is the rest mass of the free electron),\cite{Levinshtein1996} high carrier mobility ($>10^6$ cm$^2$/(Vs) in 2D electron gases (2DEGs) in InAs quantum wells,\cite{InAs:QWs_on_InP,Dempsey2025} $6500 - 8000$ cm$^2$/(Vs) in nanoflags (NFs)),\cite{Yan2025} and a relatively large Landé $g$-factor ($|g^*| = 17.0 \pm 0.5$ in InAs 2DEGs,\cite{InAs:g-factor_determination} larger than for InAs bulk, where $|g^*_{\text{bulk}}| = 14.7$).\cite{InAs:bulk_g-factor} Due to these properties, extensive studies have been conducted on InAs 1D and 2D nanostructures. Moreover, due to the strong spin-orbit interaction,\cite{InAs:NW_spin_orbit} InAs 2DEGs have been proposed as a platform for topological superconductivity in hybrid superconductor-semiconductor (S-Sm) systems.\cite{InAsOI:hybrid_superconducting_systems,InAs:topological_superconductivity, InAs:topological_superconductivity_platform, InAs:InAsOI_JJ_sheet_resistance}

InSb has an even narrower direct bandgap of $0.17$~eV and a lower effective mass than InAs ($m^*_{\text{InSb}} = 0.014\, m_e$),\cite{Levinshtein1996,Clowes2025} high carrier mobility (up to $3.5 \cdot 10^5$ cm$^2$/(Vs) in quantum wells,\cite{Yi2015,InSb:MBE_heterostructures} one order of magnitude less in free-standing nanostructures),\cite{InSb:high_mobility_NW, InSbNF:Isha, InSb:nanostructures} large Landé $g$-factor ($\sim 50$),\cite{InSb:g_factor,Clowes2025} and even stronger Rashba spin-orbit strength than InAs.\cite{InSb:NWs_spin_orbit} For these reasons, this material has recently been considered as a promising platform for topological applications in S-Sm systems, and gate controlled supercurrents are under study in InSb-based Josephson junctions. \cite{InSb:superconductivity_QW_JJ, InSb:NF_JJ_Sedi}

InAs$_{1-x}$Sb$_{x}$ is reported to have an even larger Landé $g$-factor than InAs or InSb.\cite{InAsSb:2DEG_superconductivity} Its band gap, the effective mass, and the spin-orbit splitting are reported to depend on the Sb relative concentration $x$  (band gap between $0.10$~eV and $0.36$~eV,  the ratio between effective mass and rest mass of the electron $m^*/m_e$ between $0.009$ and $0.023$).\cite{Levinshtein1996,III-V:band_parameters} Being an alloy of two promising topological materials with tunable properties, it does not surprise that this material has  recently been considered as a platform for topological superconductivity, as well.\cite{InAsSb:2DEG_superconductivity,Telkamp2025}

Here we report for the first time the growth of InAsSb nanoflags on InAs nanowires (NWs). After optimizing the growth conditions, asymmetric growth was achieved using reflection high-energy electron diffraction (RHEED) to orient the substrate with respect to metal-organic beams, in the same way as Verma et al.~did for InSb nanoflags.\cite{InSbNF:Isha} With this method, we were able to achieve $(2000 \pm 180)$~nm-long InAs$_{0.77}$Sb$_{0.23}$ NFs, $(640 \pm 50)$~nm-wide and $(130 \pm 30)$~nm-thick. Their size allowed the fabrication of Hall bar devices with standard length-to-width ratio between contacts, on which low temperature magneto-resistance measurements have been performed. From these, we derive the behavior of conductance, field-effect mobility, carrier density, and Hall mobility as a function of back-gate voltage. We report a field-effect mobility of $2.2 \cdot 10^4$ cm$^2$/(Vs) at $2.7$ K and a Hall mobility of $2.6 \cdot 10^4$ cm$^2$/(Vs) at 0.44~K, comparable to the best performing InAs and InSb nanoflags. Furthermore, we estimate the Landé $g$-factor of these InAsSb NFs to be $|g^*| = 58.7 \pm 4.0$, which is in good agreement with the recently reported value of $|g^*| \sim 55$ for an InAsSb shallow QW.\cite{InAsSb:2DEG_superconductivity}

\section{Experimental Details}\label{sec:exp}
	
The synthesis of the InAs-InAsSb heterostructures has been carried out by Au-assisted vapor-liquid-solid (VLS) growth in a chemical beam epitaxy (CBE) system (Riber Compact-21) on InAs(111)B substrates. We employed trimethylindium (TMIn), tert-buthylarsine (TBAs), and tris(dimethyamido)antimony (TDMASb) as metal-organic precursors. The growth temperature is measured with a pyrometer and the manipulator thermocouple, with an overall accuracy of $\pm 10$~°C.

The morphology of the grown samples has been studied through field-emission scanning electron microscopy (SEM), employing a Zeiss Merlin SEM with an accelerating voltage of $5$ kV and probe current between $90$ pA and $100$ pA, detecting secondary electrons. To characterize the heterostructure crystal structure and chemical composition, the InAsSb NFs have been mechanically transferred on a Cu grid and studied by transmission electron microscopy (TEM) in a JEOL JEM-F200 operated at 200~kV, equipped with an EDX spectrometer.
	
To characterize the electronic transport properties of the InAsSb NFs, they have been processed into Hall bar devices and studied at low temperature in two different cryostats. For this purpose, the samples have been sonicated in isopropyl alcohol for 35 min, to detach the NFs from the substrate and stop them from sticking to each other. 12 depositions of the solution on a Si/SiO$_2$ substrate were needed to deposit a sufficient amount of isolated NFs. Then, source-drain and four-probe Au-Ti contacts have been fabricated on top of the NFs via electron beam lithography, while the substrate acts as a global back-gate. All data presented here were measured on the same device, employing a nanoflag $(2740 \pm 20)$~nm long, $(650 \pm 10)$~nm wide, and $(94 \pm 10)$~nm thick. The distance between source and drain contacts is $(1830 \pm 15)$~nm, while the distance between probe contacts is $(900 \pm 10)$~nm in the longitudinal and $(300 \pm 10)$~nm in the lateral direction. Consistent data were obtained over several cooldowns, at 2.7 K, 1.6 K (not shown here), and 0.44 K.

\section{Results and Discussion}

The growth method used resembles the one implemented by Verma et al.\cite{InSbNF:Isha, InSbNF:understanding_morphology} First, poly-L-lysine is dropcast onto an InAs substrate and left for $30$ s, which is then rinsed in water and dried with $\text{N}_2$. A commercial water solution of gold colloidal nanoparticles (NPs) of $20$~nm in diameter (from BBI Solutions) is diluted in de-ionized (DI) water 1:8. The gold NPs, which act as seeds to catalyze VLS growth, are deposited onto the substrate, which is rinsed and blown dry again with nitrogen.

The growth process starts with the growth of InAs NW stems. We grow InAs stems for 45 min with line pressures of TMIn and TBAs of $0.35$ and $1.3$ Torr, respectively, at $390 \pm 10$°C (pyrometer reading). Afterward, the thermocouple-read temperature is raised by $\Delta T = 50$°C (to $415 \pm 10$°C pyrometer reading) while continuing stem growth for another $3$ min. This resulted in wurtzite InAs NWs with six equivalent \{112\} side facets and a length of about $1700$~nm.

Once we found the optimal conditions for InAs stem growth, instead of starting directly with the directional growth of InAsSb NFs, we started by growing InAsSb NW segments on top of the InAs stems (i.e., keeping the substrate rotating while growing), analyzing how growth parameters influence both axial and radial growth. This reduced the unnecessary complexity that the directionality of the precursor beams would have introduced in this phase. In order to obtain pure defect-free zincblende (ZB) InAsSb NWs, we have explored different line pressure ratios $r$ of group V precursors (TBAs/TDMASb). This has allowed to control Sb incorporation, and thus crystal structure. As reported in Table~\ref{tab:AS/Sb_ratio}, there is a relation between the relative atomic concentration of antimony $x$ (InAs$_{1-x}$Sb$_{x}$) and the presence of defects. For $x > 10\%$, no defects are observed. This observation is consistent with previous works on InAsSb NWs: in particular, for MBE-grown InAsSb NWs, an Sb concentration $x \sim 15\%$ is reported as threshold for the suppression of wurtzite stacking defects \cite{InAsSb:MBE_Sb_defects}. Moreover, as suggested by Ruhstorfer et al.,\cite{InAsSb:MOCVD_Sb_defects} this threshold might depend on growth dynamics, and thus on the particular process employed. Due to this effect, it has been shown that by controlling the incorporation of Sb it is possible to control the crystal phase, from the dominant wurtzite of pure InAs towards the ZB crystal structure of pure InSb.\cite{InAsSb:MBE_Sb_phase_control}

\begin{table}[t!]
	\small
	\caption{Group V line pressures (TDMASb and TBAs), their ratio $r$ (p\textsubscript{TBAs}/p\textsubscript{TDMASb}), the measured atomic composition $x$ (InAs$_{1-x}$Sb$_{x}$), and if defects are present or not. All InAs stems of these samples were grown for 48 minutes. All InAsSb NW segments were grown for 45 minutes using $0.35$ Torr of TMIn, with $\Delta T = +50$°C, except for $r$ = 5.5, grown at $\Delta T = +30$°C.}
	\label{tab:AS/Sb_ratio}
	\begin{tabular*}{0.48\textwidth}{@{\extracolsep{\fill}}lllll}
		\hline
		p\textsubscript{TDMASb} [Torr] & p\textsubscript{TBAs} [Torr] & $r$ & $x$ & Crystal quality \\
		\hline
		0.2 & 1.1 & 5.5 & $6 \%$ & Defects \\
		0.3 & 1.2 & 4 & $7.5 \%$ & Defects \\
		0.4 & 1.1 & 2.75 & $7.7 \%$ & Defects \\
		0.5 & 1.0 & 2 & $13.3 \%$ & pure ZB \\
		0.6 & 0.7 & 1.17 & $10.2 \%$ & pure ZB \\
		0.6 & 0.9 & 1.5 & $15.6 \%$ & pure ZB \\
		\hline
	\end{tabular*}
\end{table}

Having found the best conditions to grow InAsSb NWs, we have applied these conditions to grow InAsSb NFs using the \textit{directional growth} method. For this purpose, at the end of the growth of the InAs stems, substrate rotation is stopped, and, using RHEED, the substrate is oriented such that one of the six equivalent \{112\} sidewalls of the  InAs stems faces the TBAs beam. To understand the geometry of the orientation process, we refer to Section S1 of the Electronic Supplementary Information (ESI). Finally, the growth of InAsSb NFs is initiated. The aim of our investigation has been to grow samples with an acceptable yield of defect-free InAsSb NFs thinner than $100$~nm and wider than $600$~nm. 

The main limit of Sb incorporation is the low vapor pressure of TDMASb at room temperature, which limits the maximum line pressure to $0.6$ Torr for a stable reading. Besides, once the TDMASb line pressure is fixed to $0.6$ Torr, reducing the TBAs line pressure below $0.9$ Torr significantly reduces the lateral growth rate of the NFs. Thus, we first decided to grow InAsSb NFs using line pressures of $0.9$ Torr and $0.6$ Torr of TBAs and TDMASb, respectively ($r = 1.5$). Since we also found that increasing TMIn line pressure increases the radial growth rate,\cite{Serra2025} the TMIn line pressure during growth was increased from $0.35$ Torr (for NW growth) to $0.7$ Torr (for NF growth).

\begin{figure}[t!]
	\centering
	\includegraphics[width=0.62\columnwidth]{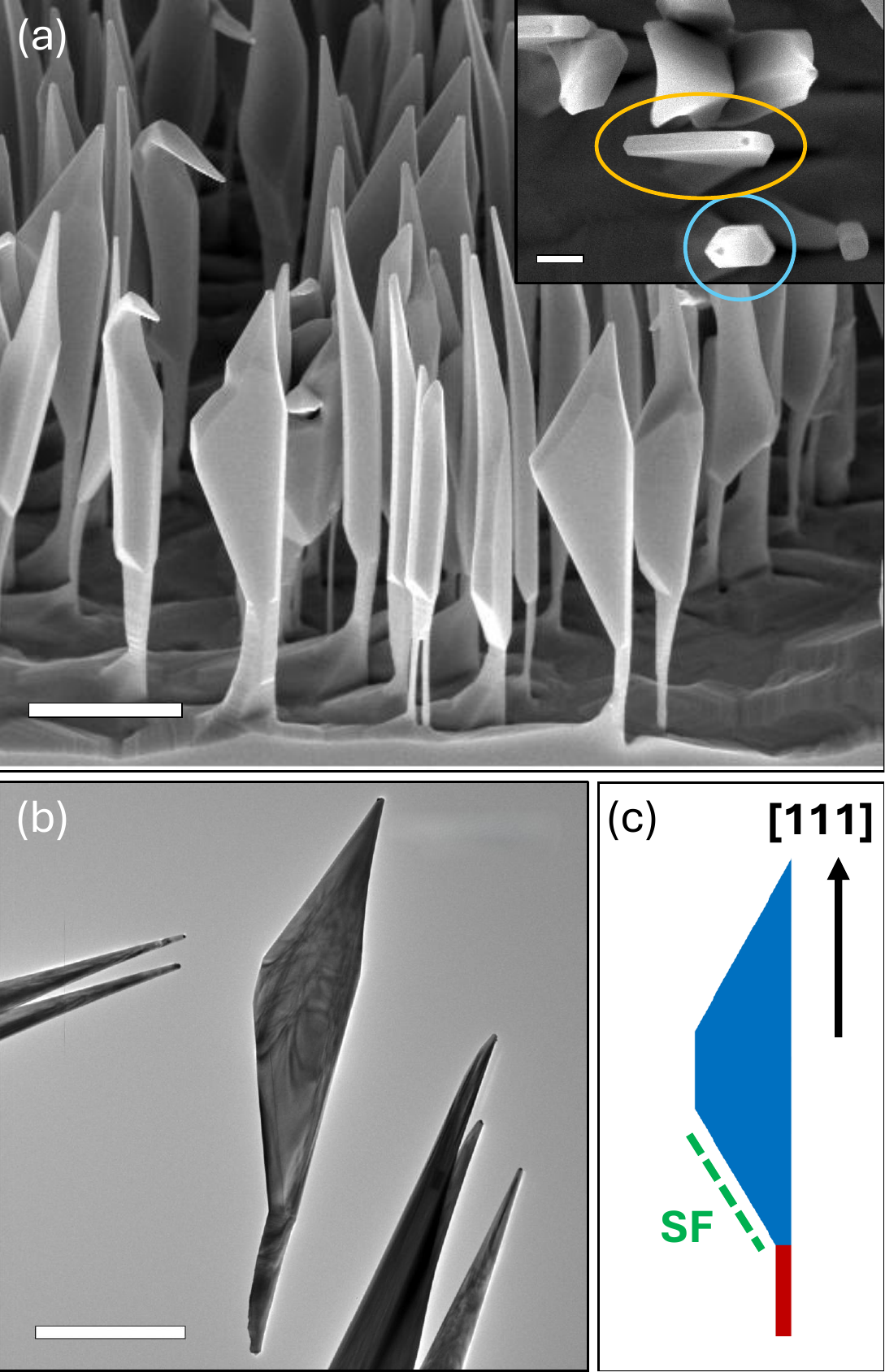}
	\caption{(a) 45° tilted-view SEM image of InAsSb NFs under optimized growth conditions. Length marker $1$ \textmu m. Inset: top-view SEM image of the same sample, highlighting in yellow a NF with the desired morphology, in blue a much narrower and thicker NF that grows in the opposite direction. Length marker $200$~nm. (b) TEM micrograph of an isolated InAsSb NF. A single stacking fault originates at the stem-NF interface and propagates at the edge of the NF. Length marker $1$ \textmu m. (c) Sketch of a NF: in red the InAs stem, in blue the InAsSb NF, in green the stacking fault (SF).}
	\label{fgr:panel1}
\end{figure}

In fact, an important issue is that the directional growth method produces two different kinds of NFs:
\begin{itemize}[leftmargin=*]
	\item thin and wide NFs, which is the desired morphology. They grow in the direction of the incoming TBAs beam (i.e., towards it). One example is circled in yellow in the inset to Fig.~\ref{fgr:panel1}a. While they grow virtually without defects, their yield is low ($5.4 \%$), and about $10 \%$ of them are wider than $600$~nm and thinner than $100$~nm.
	\item much narrower and thicker NFs. They grow in the opposite direction (i.e., away from the TBAs beam). One example is circled in blue in the inset to Fig.~\ref{fgr:panel1}a. Their presence reduces the yield of NFs with the desired morphology.
\end{itemize}

\begin{figure}[ht]
	\centering
	\includegraphics[width=\columnwidth]{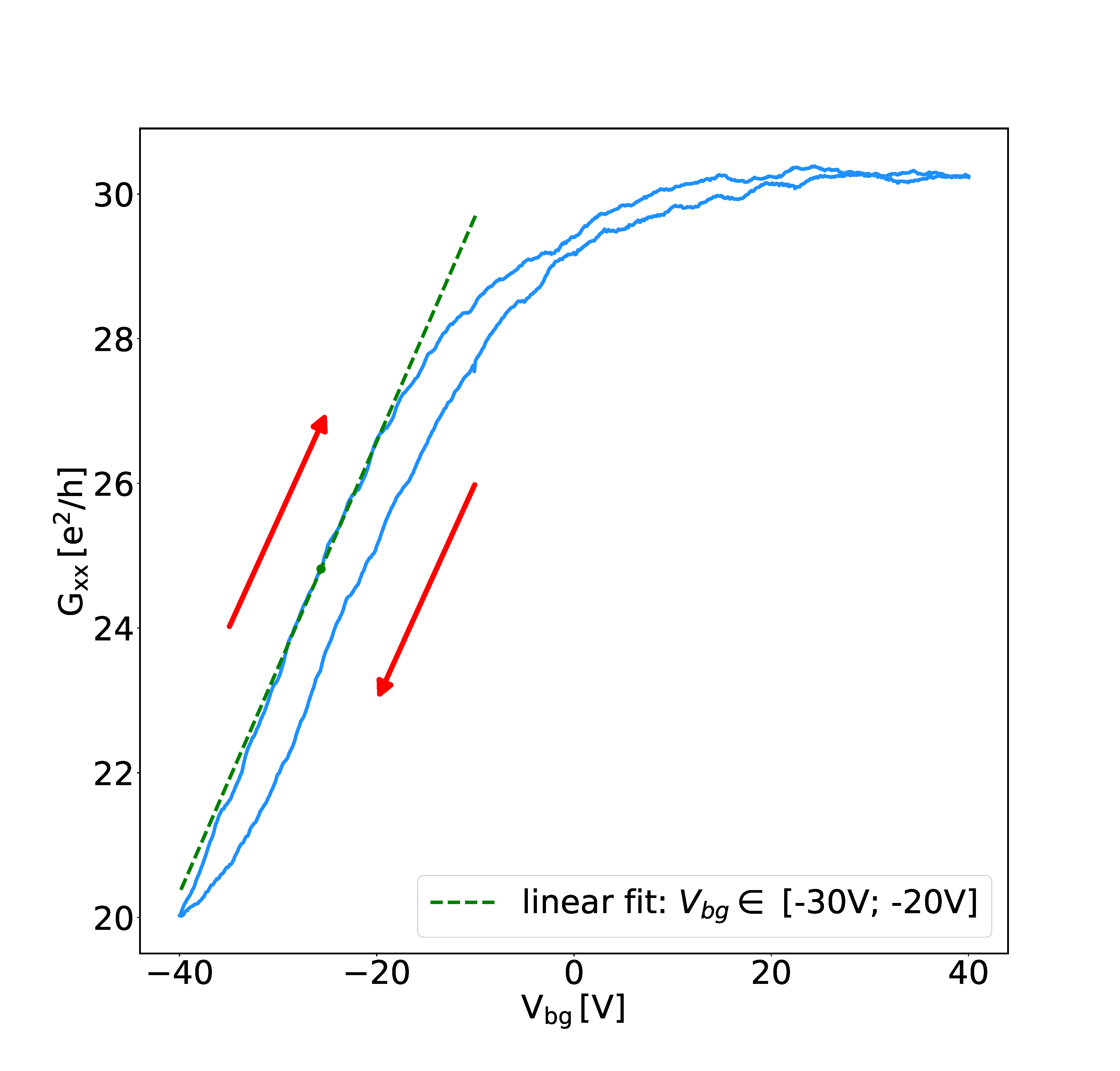}
	\caption{Four-probe longitudinal conductance $G_{xx}$ as a function of back-gate voltage $V_{bg}$, for two sweep directions, forward and backward, as indicated by the red arrows. In green, a linear fit to the forward-sweep curve in the [$-30$ V; $-20$ V] range, to obtain field-effect mobility. $T = 2.7$~K. Sweep rate: $0.086$~V/s.}
	\label{fgr:panel2}
\end{figure}

After experimenting with different growth recipes and growth parameters, we found that the best way to increase the yield is to increase TDMASb line pressure over $0.6$ Torr. In order to do so, the TDMASb bubbler has been heated to $30$°C, allowing to reach a stable reading of $0.9$ Torr. This way, it has been possible to grow the NFs with $0.7$ Torr, $1.1$ Torr, and $0.9$ Torr of TMIn, TBAs, and TDMASb, respectively, and obtain an acceptable yield of $5.7 \%$ of InAsSb NFs with the desired morphology. Among these, $15 - 20 \%$ were wider than $600$~nm and thinner than $100$~nm. We are confident that with future optimization, a better size and shape control can be achieved, as well as a higher yield.

A $45$°-tilted SEM image of a sample obtained under these optimized growth conditions is reported in Fig.~\ref{fgr:panel1}a (top-view SEM image in the inset). The image shows InAsSb NFs with an average size of length $(2000 \pm 180)$~nm, width $(640 \pm 50)$~nm, and thickness $(130 \pm 30)$~nm. The TEM micrograph in Fig.~\ref{fgr:panel1}b displays one of the InAsSb NFs: no defects are present, except for a single planar defect originating at the interface between InAs stem and NF, which runs along the bottom lateral side of the NF. Consistent results were obtained from a total of 8 NFs. The occurrence of this defect is consistent with what has been reported for MBE-grown InSb NFs.\cite{InSb:nanosails} From EDX measurements we determined that NFs grown this way have an average composition of $53 \%$ indium, $36 \%$ arsenic, and $11 \%$ antimony (each value with a $\pm 5 \%$ error), resulting in InAs$_{0.77}$Sb$_{0.23}$ NFs.

\begin{figure*}[ht]
	\centering
	\includegraphics[width=\textwidth]{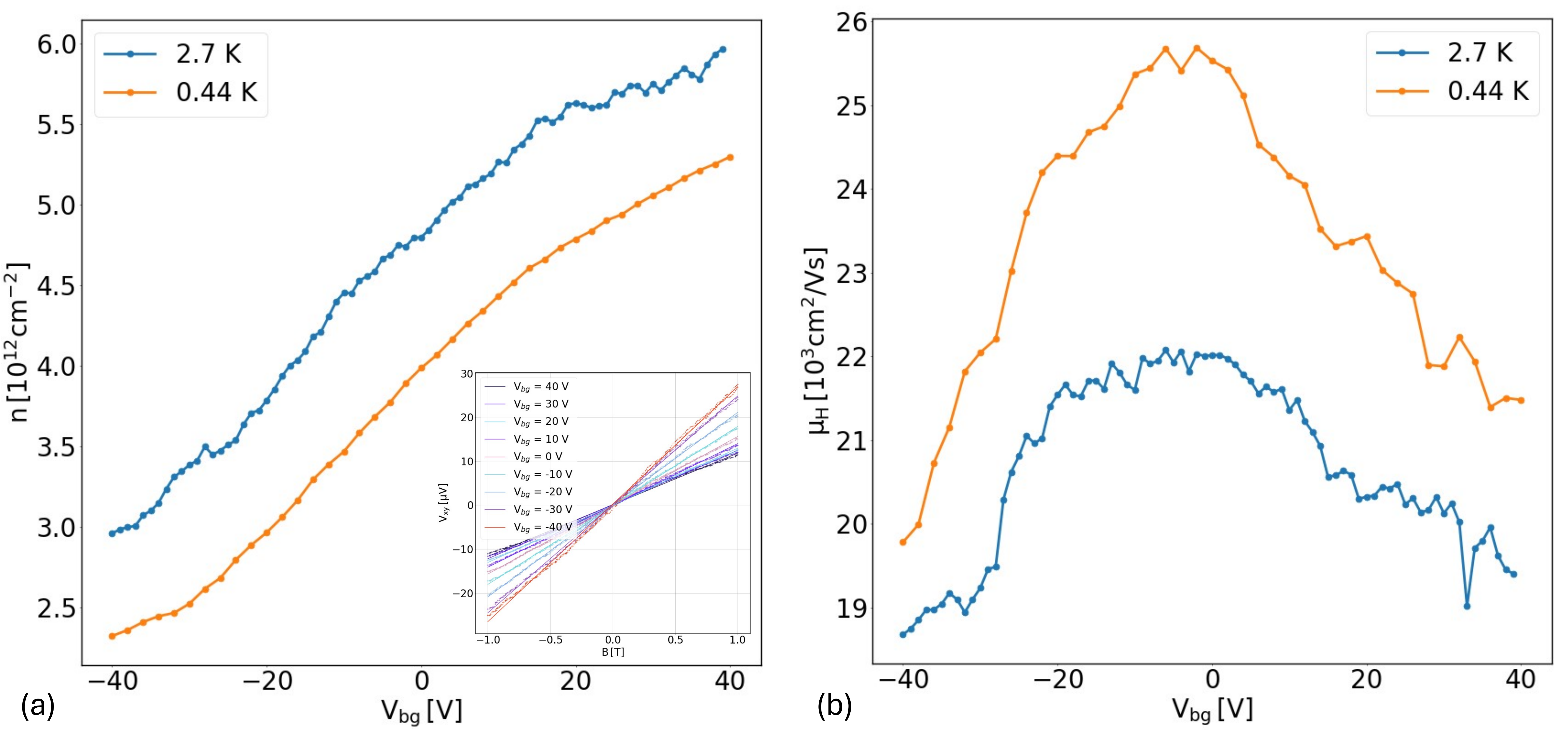}
	\caption{Hall effect measurements at two different temperatures, $2.7$~K and $0.44$~K. (a) Charge carrier density $n$ as a function of back-gate voltage $V_{bg}$. In the inset, the four-probe transverse voltage $V_{xy}$ is shown as a function of magnetic field $B$ at $0.44$~K for different back-gate voltages. The data has been corrected for the misalignment of the voltage probes in the longitudinal direction. (b) Hall mobility $\mu_H$ as a function of back-gate voltage.} 
	\label{fgr:panel3}
\end{figure*}

Low temperature transport measurements have been performed at $2.7$ K, at $B = 0$, $B = 0.25$~T, and $B = -0.25$~T. First, we measured source-drain current $I_{sd}$ and voltage $V_{sd}$, and four-probe longitudinal ($V_{xx}$) and transverse ($V_{xy}$) voltages as a function of back-gate voltage. For details, see Section S2 of the ESI. The data was measured with an AC current bias of $100$ nA and performing back-gate voltage sweeps, back and forth, between $\pm 40$ V. From the measurements at $B = 0$, we calculated the longitudinal conductance $G_{xx}$ as follows:
\begin{equation}
	G_{xx} = \frac{R_{xx}}{R^2_{xx} + R^2_{xy}} \approx \frac{1}{R_{xx}},
\end{equation}
where $R_{xx} = V_{xx}/I_{sd}$, and $R_{xy} = V_{xy}/I_{sd}$ should be $\sim 0$ $\Omega$ at zero magnetic field. Experimentally, we found $R_{xy} \sim 15-20\, \Omega$, which we attribute to a misalignment of lateral probes in the longitudinal direction (probes 2 and 3 in Fig.~S2 in the ESI). By neglecting this term, since $R^2_{xx} >> R^2_{xy}$, we introduce a relative error of less than $1 \%$ in the denominator.

Fig.~\ref{fgr:panel2} shows the behavior of $G_{xx}$ as a function of back-gate voltage. The plot shows an increasing $G_{xx}$ for increasing back-gate voltages. This means that InAsSb NFs behave as n-type semiconductor, i.e., electrons are the majority carriers. Moreover, by looking at the behavior of $G_{xx}$ at $-40$ V it is evident that the channel stays open even for negative gate voltages, as conductance never goes to zero, but remains above $20\, e^2/h$. This is consistent with what has been reported for InAsSb NWs with ZB structure at intermediate concentrations of Sb,\cite{Sestof:no_pinch-off} and with the observed behavior of $I_{sd}$ versus $V_{bg}$, which does not show a current pinch-off (reported in Section S3 of the ESI). Such features are compatible with the presence of a surface conduction channel due to Fermi level pinning in the conduction band at the surface, a phenomenon already reported for InAs in e.g.~InAs/InAlAs heterostructures\cite{InAs/InAlAs:multiple_channels} or InAs NFs.\cite{Seidl2019} For a deeper discussion of the presence of two conduction channels and its implications see Section S4 of the ESI.

Another feature of the conductance curves is  a noticeable hysteresis. We attribute this to charge traps at the semiconductor-gate interface, which alter the effective gate field seen by the conduction channel. This behavior is in line with what has been reported for InAs NWs,\cite{InAs:hysteresis} where it is found to be strongly dependent on sweep rate and range.

Always in Fig.~\ref{fgr:panel2}, we display the result of a linear fit to the raw data for $V_{bg} \in \left[-30\, V; -20\, V\right]$. From the slope of this curve we can estimate the field-effect mobility as:
\begin{equation}\label{eq:field-effect_mobility}
	\mu_{FE} = \frac{L}{W C_{ox}} \cdot \frac{\partial G_{xx}}{\partial V_{bg}},
\end{equation}
where $L = 900$~nm and $W = 300$~nm are the distances between the voltage probes in the longitudinal and transversal direction, respectively, and $C_{ox}$ is the capacitance per unit area of the thin SiO$_2$ oxide layer. Given the SiO$_2$ layer thickness of $t_{ox} = 285$~nm and 
\begin{equation}
	C_{ox} = \frac{\varepsilon_{\text{SiO}_2}\varepsilon_0}{t_{ox}},
\end{equation}
we found the oxide capacitance to be $C_{ox} = 1.2 \cdot 10^{-8}$ F/cm$^2$. This value is in good agreement with previous work.\cite{InSbNF:Isha} The field-effect mobility obtained from the linear fit to the conductance curve is $\mu_{FE}= 2.2 \cdot 10^4$ cm$^2$/(Vs) (at $2.7$ K).

Hall effect measurements were taken, for different $V_{bg}$ values, by measuring source-drain current $I_{sd}$, longitudinal four-probe voltage drop $V_{xx}$, and Hall voltage $V_{xy}$, with lock-in amplifiers with an AC current bias of $100$ nA. Since the channel is always open, we explored $V_{bg}$ values in the whole $\pm 40\,V$ range. In particular, the following measurements were taken: at $2.7$ K, for $B = 0$, $+0.25$~T, and $-0.25$~T; at $0.44$~K, with $B$-field sweeps in the [$-1$~T; $+1$~T] range. At $0.44$~K, we also acquired data for back-gate voltages between $\pm$ 40 V, changing the magnetic field from $8$~T to $-8$~T.

The inset to Fig.~\ref{fgr:panel3}a shows the behavior of $V_{xy}$ as a function of magnetic field, for some of the $V_{bg}$ values explored. These curves allow to calculate the charge carrier density $n$ and the Hall mobility $\mu_H$ as follows:
\begin{equation} \label{eq:carrier_density}
	n = \frac{I_{sd}}{e} \cdot \left(\frac{\partial V_{xy}}{\partial B}\right)^{-1},
\end{equation}
\begin{equation} \label{eq:hall_mobility}
	\mu_H = \frac{L}{W \cdot \langle V_{xx} \rangle} \cdot \left(\frac{\partial V_{xy}}{\partial B}\right),
\end{equation}
where $I_{sd}$ = 100 nA is the bias current, $e$ the electron charge, and $\langle V_{xx} \rangle$ the longitudinal four-probe voltage at $B = 0$. Both charge carrier density and Hall mobility at $2.7$~K and $0.44$~K are represented against back-gate voltage in Fig.~\ref{fgr:panel3}a and b, respectively. Increasing back-gate voltage increases charge carrier density, which is consistent with the behavior of an n-type semiconductor. For what concerns temperature, electrons in the NFs behave as expected for particles obeying the Fermi-Dirac statistics: as $T$ increases, more carriers become available for transport. With $\lambda_F = 1 / (2 \pi n)$ we can calculate the Fermi wavelength corresponding to these carrier concentrations and obtain $\lambda_F$ between $16$ and $23$~nm, which  places these nanoflags in the quasi-2D transport regime (considering their thickness of $94$~nm).

While at $2.7$ K peak Hall mobility is equal to the peak field-effect mobility at the same temperature (cf.~Fig.~\ref{fgr:panel2}), at $0.44$~K Hall mobility has a peak of about $2.6 \cdot 10^4$ cm$^2$/(Vs) (at a corresponding carrier density $n \approx 4 \cdot 10^{12}$~cm$^{-2}$, see Fig.~\ref{fgr:panel3}a), which is in line with Hall mobilities reported in previous studies for InSb nanoflags,\cite{InSbNF:Isha, InSb:nanosheets, InSb:nanosails, InSb:nanostructures} InSb shallow quantum wells,\cite{InSb:QW_proximity_effect} and InAsSb shallow QWs.\cite{InAsSb:2DEG_superconductivity} To pinpoint the scattering mechanisms that limit mobility (like surface or interface scattering and the role of the stacking fault) would require measurements over a wider temperature range, which is left for future work.

\begin{figure}[t!]
	\centering
	\includegraphics[angle=0, width=\columnwidth]{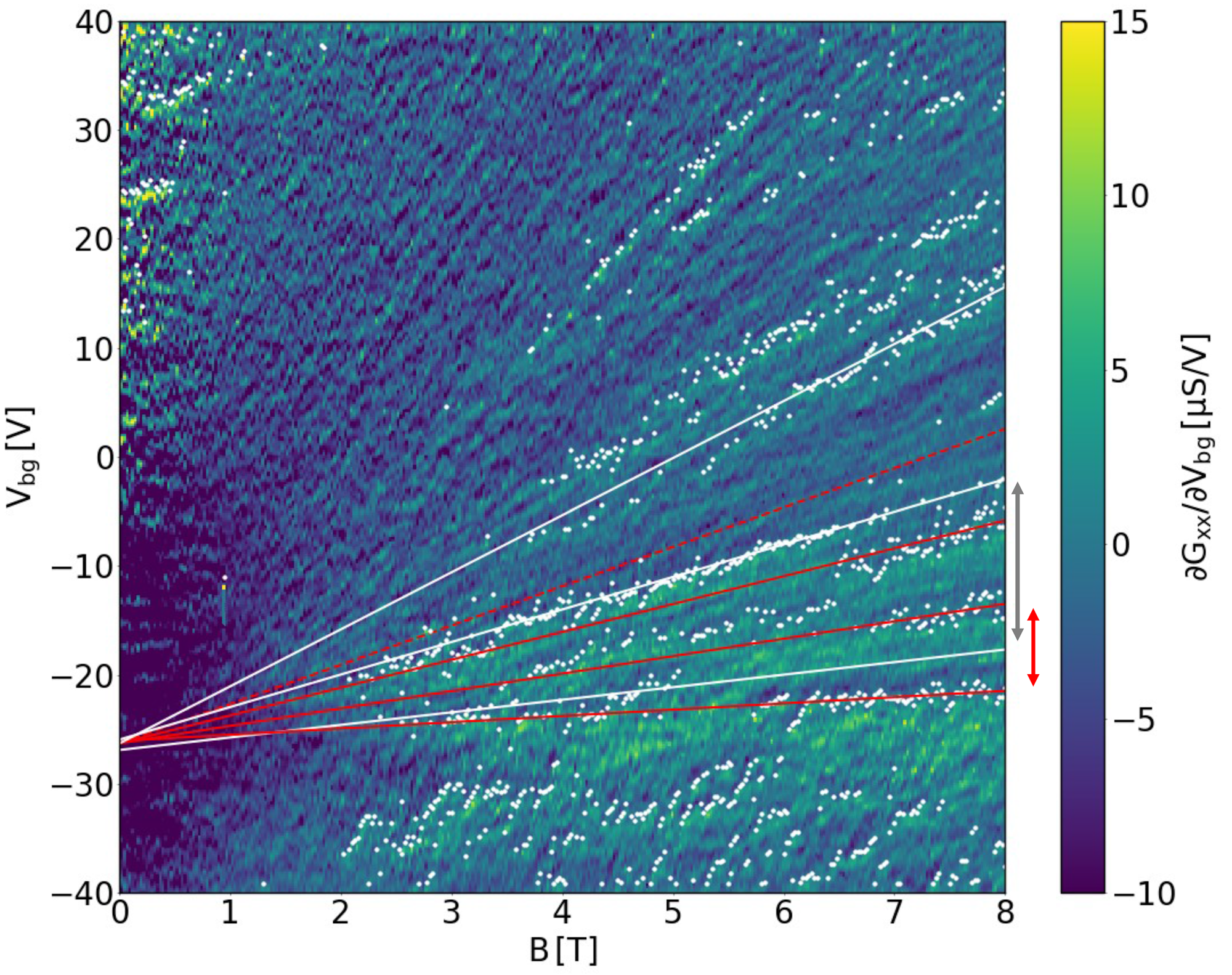}
	\caption{Colormap of the variation in longitudinal conductance $\partial G_{xx} / \partial V_{bg}$ with back-gate voltage, as a function of back-gate voltage $V_{bg}$ and magnetic field $B$. The relative minima of $G_{xx}$ along $V_{bg}$ for different values of $B$ are superimposed as white dots. Linear fits to subsets of the dots are represented in white for Landau levels with $N = 0,1,2$. The red continuous lines represent linear fits corresponding to the Zeeman-split Landau levels $N = 0$ and $N = 1$. On the right, a gray arrow indicates the Landau level splitting $\Delta E_{LL}$ at $B = 8$~T, while the red arrow indicates the Zeeman splitting $\Delta E_Z$ of the LLs, at the same magnetic field.}
	\label{fgr:panel4}
\end{figure}

Finally, from the measurements at $0.44$~K, taken while performing $V_{bg}$ sweeps from $-40$ V to $+40$ V at values of magnetic field ranging from $0$ to $8$~T, a detailed colormap of the four-probe longitudinal conductance variation $\partial G_{xx} / \partial V_{bg}$ with respect to back-gate voltage has been obtained and is shown in Fig.~\ref{fgr:panel4}. The derivative of $G_{xx}$ has been taken to highlight magneto-conductance effects and subtract the linear behavior from the conductance. Data taken for negative magnetic fields are consistent with this map. 

From the theory of magneto-transport in 2DEGs,\cite{Yu2005,Ihn:magnetotransport_2D} we know that if Shubnikov-de-Haas oscillations are present, conductance shows local minima, aligned on lines,\cite{Turini:InSbJJ} each one corresponding to a different integer filling factor $\nu$ of Landau levels (LLs), since the energy of LLs depends linearly on the magnetic field.

Before searching for the minima, the longitudinal conductance has been smoothed with a Savitsky-Golay filter of order 1 and a window of 9 data points, using \texttt{signal.savgol\_filter} from the \textit{scipy} library. Then, using \texttt{signal.argrelmin} from the same library (order 11), the relative minima of $G_{xx}$ were located and are shown in Fig.~\ref{fgr:panel4} as white dots over the colormap of the conductance variation.

In the range $B \lesssim 5$~T, three sets of minima were found, one with slope $k_0 = 1.2 \pm 0.1$~V/T, the second with slope $k_1 = 2.99 \pm 0.03$~V/T, the third with slope $k_2 = 5.24 \pm 0.01$~V/T. The best-fitting lines through them are colored in white in Fig.~\ref{fgr:panel4}. The lines correspond to filling factors $\nu = 2, 4, 6$, respectively, indicating the three lowest energy spin-degenerate Landau levels, with $N = 0, 1, 2$. The slope of these lines is related to the quantum number $N$ as
\begin{equation}
	k_N = \frac{e}{\alpha \pi \hslash} \left( N + \frac{1}{2} \right) ,
\end{equation}
with $\alpha$ the back-gate lever arm, i.e., the ratio between induced charge $n$ and applied back-gate voltage $V_{bg}$. The experimental results agree with the expected ratio between the slopes, $k_1/k_0 = 3$ and $k_2/k_1 = 5/3$. For $B \gtrsim 5$~T, the first white line splits into two lines in a symmetric fashion (shown in red Fig.~\ref{fgr:panel4}). This indicates that the spin-degeneracy is lifted by the Zeeman splitting, which is expected in systems with large Landé $g$-factor like InAsSb.

The energy gap due to the lifting of degeneracy is
\begin{equation}
	\Delta E_Z = g^* \mu_B B = g^* \frac{e \hslash}{2 m_e} B
\end{equation}
and  the energy difference between LLs is
\begin{equation}
	\Delta E_{LL} = \hslash \omega_c = \frac{\hslash e B}{m^*}.
\end{equation}
It is therefore possible to calculate the Landé $g$-factor as:
\begin{equation} \label{eq:effective_g-factor}
	|g^*| = \frac{2}{m^*/m_e} \frac{\Delta E_Z}{\Delta E_{LL}}.
\end{equation}
At $B = 8$~T, the average separation between the white lines is $\delta E_{LL} = \left( 16.6 \pm 1.0 \right)$~V, while $\delta E_Z = \left( 7.60 \pm 0.05 \right)$ V. Further details on this analysis are reported in Section S5 of the ESI.

In order to obtain $|g^*|$, the effective mass of InAs$_{0.77}$Sb$_{0.23}$ needs to be known. Since the effective masses of InAs and InSb are $m^*_{\text{InAs}} = 0.023\, m_e$ and $m^*_{\text{InSb}} = 0.014\, m_e$,\cite{Levinshtein1996} an estimate of the effective mass in InAs$_{0.77}$Sb$_{0.23}$ can be found by considering the reported dependence of InAs$_{1-x}$Sb$_{x}$ effective mass on $x$ in the bulk: $m^*/m_e = 0.023 - 0.039 \, x + 0.03 \, x^2$.\cite{Levinshtein1996,Rogalski1989,Rogalski1989a} Here, for $x = 0.23$ we find $m^* = 0.0156 \, m_e$, which is close to the value for InSb. This is due to the bowing in the dependence of effective mass with respect to $x$.\cite{InAsSb:band_gap,III-V:band_parameters,Levinshtein1996,Rogalski1989,Rogalski1989a} For further details, see Section S6 of the ESI. This interpolation procedure provides a reasonable estimate of the effective mass of these NFs. A direct measurement, for example from the temperature dependence of Shubnikov-de Haas oscillations,\cite{Telkamp2025} is left for future work.

Plugging these numbers into Eq.~\ref{eq:effective_g-factor}, we estimate the Landé $g$-factor of InAs$_{0.77}$Sb$_{0.23}$ NFs to be $|g^*| = 58.7 \pm 4.0$, which is higher than the previously reported value for InSb NFs.\cite{Turini:InSbJJ}

This Landé $g$-factor is in good agreement with the recently reported value of $|g^*| \sim 55$ for an InAsSb shallow QW.\cite{InAsSb:2DEG_superconductivity} A slightly larger value for the NFs is expected here, since the reported QW had a lower Sb concentration and thus a larger effective mass.

To put our results in perspective, we close with some general considerations. An advantage of the InAsSb material system compared to InAs and InSb is the larger Landé $g$-factor while maintaining a comparable mobility. Ternary compounds are generally more challenging to grow than binaries, but on the other hand, they allow control on lattice constant, band gap, and effective mass via composition engineering. Compared to one-dimensional nanowires, two-dimensional systems allow greater flexibility in device design and fabrication. Finally, compared to NFs, quantum wells constitute a scalable technology that can provide high mobility, but their active region is typically buried more than $100$~nm under the surface, which makes proximitization by a superconductor deposited on the surface of the semiconductor less effective. The proximity can be improved in shallow quantum wells, which are grown closer to the surface, however at the expense of a reduced mobility.

\section{Conclusions}

In conclusion, we successfully synthesized, for the first time, InAs$_{0.77}$Sb$_{0.23}$ NFs of high crystal quality that in average are $(2000 \pm 180)$~nm long, $(640 \pm 50)$~nm wide, and $(130 \pm 30)$~nm thick. The NFs were obtained using a directional growth method previously employed by Verma et al.~to grow InSb NFs.\cite{InSbNF:Isha} We found that these NFs display mobility up to $2.6 \cdot 10^4$ cm$^2$/(Vs) at $0.44$~K, which is comparable to the best performing InSb and InAs nanoflags. InAs$_{0.77}$Sb$_{0.23}$ NFs have a large Landé $g$-factor of $58.7 \pm 4.0$ and show signs of surface conduction. These characteristics make them a promising platform to study topological superconductivity in hybrid S-Sm systems and a good candidate to host exotic bound states.

\section*{Author contributions}


Sebastian Serra: investigation, data curation, visualization, software, and writing – original draft. Gaurav Shukla: resources, writing – review \& editing. Giada Bucci: resources, writing – review \& editing. Robert Sorodoc: resources, writing – review \& editing. Valentina Zannier: resources, writing – review \& editing. Fabio Beltram: resources, writing – review \& editing. Lucia Sorba: conceptualization, methodology, resources, validation, supervision, writing – original draft. Stefan Heun: investigation, validation, supervision, writing – original draft.
	
\section*{Conflicts of interest}

There are no conflicts to declare.

\section*{Data availability}
	
The data supporting this article have been included as part of the Supplementary Information.

\section*{Acknowledgements}
	
The authors acknowledge support from project PNRR MUR Project No.~PE0000023-NQSTI and from PRIN2022 2022-PH852L(PE3) TopoFlags-“Non-reciprocal supercurrent and topological transitions in hybrid Nb-InSb nanoflags” funded by the European community-Next Generation EU within the program “PNRR Missione 4-Componente 2-Investimento 1.1 Fondo per il Programma Nazionale di Ricerca e Progetti di Rilevante Interesse Nazionale (PRIN)”. We acknowledge the Center for Instrument Sharing of the University of Pisa (CISUP) for the TEM facility.
	
	
\balance
	
	
\bibliography{references} 
\bibliographystyle{rsc} 

\end{document}